\documentclass[11pt,superscriptaddress]{article}

\usepackage[a4paper, total={5.8in, 9in}]{geometry}
\usepackage[indent,skip=4pt]{parskip}
\usepackage{amsmath,amssymb,amsfonts,amsthm,mathrsfs,mathtools,cases}
\usepackage{dsfont}
\usepackage{authblk}
\usepackage[colorlinks=true,linkcolor=blue, citecolor=blue, bookmarks]{hyperref}
\usepackage{cleveref}
\usepackage{microtype}
\usepackage{comment}
\usepackage{graphicx,adjustbox} 
\usepackage{subfig}
\usepackage{wrapfig}
\usepackage[font=small,labelfont=bf]{caption}
\usepackage[dvipsnames]{xcolor}  
\usepackage{booktabs}
\usepackage{braket}
\usepackage[nottoc]{tocbibind}
\usepackage{nicematrix}
\usepackage{tikz}
\usepackage{pgfplots}
\usepackage{cite}
\pgfplotsset{compat=1.18}
\usepackage{enumitem}

\usetikzlibrary{matrix}
\usetikzlibrary{decorations.markings}
\usetikzlibrary{patterns}
\usetikzlibrary{arrows.meta}

\numberwithin{equation}{section}

\crefname{figure}{Fig.}{Figs.}
\crefname{equation}{Eq.}{Eqs.}
\crefname{section}{Sec.}{Secs.}
\crefname{appendix}{Appendix}{Appendices}
  
\colorlet{darkerblue}{MidnightBlue!20!black}
\colorlet{lightblue}{blue!70!white}

\hypersetup{
    colorlinks,
    linkcolor={Blue},
    citecolor={Blue},
    urlcolor={Blue},
}

\newcommand{\beq}{\begin{equation}}
\newcommand{\eeq}{\end{equation}}
\newcommand{\beqa}{\begin{eqnarray}}
\newcommand{\eeqa}{\end{eqnarray}}

\newcommand{\Tr}{\operatorname{Tr}}

\usepackage[framemethod=TikZ]{mdframed}

\title{\textbf{A Gaussian asymmetry measure}}
\author[1]{Riccardo Travaglino}
\author[1]{Pasquale Calabrese}

\affil[1]{\textit{SISSA and INFN Sezione di Trieste, via Bonomea 265, I-34136 Trieste, Italy}}

\date{}

\begin{document}
\maketitle

\begin{abstract}
The study of Entanglement Asymmetry  has emerged in recent years as a powerful tool to characterise the symmetry properties of quantum states in relation to a given charge operator through the lens of entanglement. While extremely powerful and general, the standard definition of asymmetry introduces significant non-Gaussian features in free-fermionic systems, leading to certain analytical limitations. 
In this work, we introduce an asymmetry measure that remains strictly within the Gaussian manifold and analyse its properties. In particular, we show that it quantifies the minimal distance between a Gaussian state and the manifold of symmetric Gaussian states.
We further demonstrate that this measure captures the established dynamical signatures of entanglement asymmetry, such as the Mpemba effect, symmetry restoration, and the lack thereof.  
The Gaussian structure allows these novel asymmetry measures to be computed exactly using correlation matrix techniques, and to be described asymptotically through the quasiparticle picture. We also comment on the possibility of using charge fluctuations to characterise the asymmetry of a Gaussian state.

\end{abstract}

\maketitle 
\newpage
\section{Introduction }
The characterisation of the symmetry properties of quantum states has attracted considerable interest in recent years. While global symmetries are relatively easy to characterise, through the conserved charges of the system, and in general fix the properties of ground and equilibrium states,  
the symmetry properties of general dynamical quantum systems are much more subtle to capture. 
For example, even though the Mermin–Wagner theorem forbids spontaneous symmetry breaking of continuous symmetries in the ground state, a general quantum state $\rho$ evolving under a Hamiltonian $H$ that commutes with a $U(1)$ charge generated by an operator $Q$ can nevertheless exhibit strong symmetry breaking. This may occur either at the level of the full system, $[\rho,Q] \neq 0$, or at the level of subsystems, $[\rho_A, Q_A] \neq 0$, where the reduced density matrix is defined as $\rho_A = \operatorname{Tr}_{\overline{A}}[\rho]$.
Considering properties of the local density matrix has been of fundamental importance in the last decades, as it has allowed to understand the mechanisms underlying quantum thermalisation, mostly through the lens of entanglement \cite{Deutsch1991,srednicki1,Rigol:2007juv,DAlessio:2015qtq,Deutsch_2018,abanin2019many,Kaufman_2016}; hence it is natural to study symmetries from the same point of view, as it allows to make general thermodynamic and hydrodynamic (transport related) statements.

It is therefore of great theoretical importance to have a diagnostic tool to detect the presence or absence of a given symmetry and, in the latter case, to quantify the degree of its breaking at the level of the subsystem density matrix $\rho_A$.
While a natural first approach would be to consider charge fluctuations within $A$, this naive strategy fails, as the distribution of charge fluctuations in out-of-equilibrium systems is highly nontrivial even for symmetric states \cite{parez2021quasiparticle,bertini2023nonequilibrium}.
A more effective quantifier, termed entanglement asymmetry (EA), has been introduced in recent years \cite{Gour2009,Marvian:2014awa,Casini:2020rgj,ares2023entanglement}: focusing on a $U(1)$ symmetry with generator $Q = \sum_i q_i$, this is defined as the relative entropy between the local state $\rho_A$ and its symmetrisation $\rho_{A,Q}$ 
\begin{equation}
    \Delta S(\rho_A) = S(\rho_A||\rho_{A,Q}) =S(\rho_{A,Q}) - S(\rho_A)
\end{equation}
where $\rho_{A,Q} = \sum_q P_q \rho P_q$   is obtained by twirling over the $U(1)$ group generated by $Q_A$ (i.e., projecting onto definite local charge sectors). 
Being a relative entropy, $  \Delta S_A  $ vanishes if and only if $  [\rho_A, Q_A] = 0  $, namely if the state is locally symmetric; moreover,  it constitutes a good measure of symmetry breaking as it minimises the relative entropy of $\rho$ with any symmetric state \cite{Gour2009}. 
Reintroduced in the context of many-body quench dynamics \cite{ares2023entanglement}, entanglement asymmetry has rapidly emerged as a versatile diagnostic of nonequilibrium symmetry restoration and its absence, enabling the identification of counterintuitive relaxation phenomena, known as the quantum Mpemba effect (QME) \cite{ares2023entanglement,Ares2025qme,Teza_2026,Calabrese_2026,summer2026} (by analogy with a famous classical effect
\cite{Mpemba_1969}) according to which states which exhibit stronger symmetry breaking at initial times can restore such symmetry faster. 
Such effect has been subject of an intensive investigation, allowing to observe it in free and integrable systems \cite{murciano2024entanglement,rylands2024microscopic,ares2025simpler,Chalas_2024,Rylands_2024,shion2,Banerjee2025,yamashika2025,Gibbins_2025,ares2025mixed,Klobas_2024},
free systems with linear dissipation \cite{Caceffo_2024} or continuous Gaussian measurements \cite{digiulio2025}, quantum circuits \cite{Liu2024,turkeshi2025,ares2025rqc,Klobas2025,Foligno_2025,xu2025observation,Yu_2025,aditya2025,li2026}, random models \cite{Liu_2025,russotto2025dynamics},
long-range systems \cite{Joshi2024,Yamashika_2026b,hallam2025}, and chaotic models \cite{bhore2025,qi2025,yamashika2026,muller2026,yu2026,mcroberts2026}. 
Beyond the insight on dynamical phenomena, the EA has also been characterised in ground states \cite{Capizzi2023,Ferro_2024,Khor2024,Maric2025,ferro2026} and quantum field theories \cite{Casini_2022,Kusuki2024,Fossati2024,chen2024renyi,Lastres2025,chen2024ph,Benini_2025,Fossati2025,benini2026asymmery,GattoLamas25,benini25cat,florio2025,zhang2025}, general bounds on its value has been provided  \cite{Capizzi_2024,mazzoni2026,gotta2026} and its typical behavior in random states has been analised \cite{ares2024blackhole,Russotto2025,Russotto_2025b,Fujimura2026,Joshi2026}.

While the EA has the advantage of being extremely general, its application to the study of symmetry restoration in free-fermionic systems may appear somewhat odd. In fact, considering a fermionic Gaussian state evolving via a quadratic Hamiltonian, the evolution at all times is restricted to the manifold of Gaussian states. On the other hand,  $\rho_{A,Q}$ is a highly non-Gaussian state, meaning that the dynamics of $\rho_A$ lies within a set of states which are completely disconnected from it, except for asymptotically large times, when the the symmetry is restored together with Gaussianity (up to pathological situations \cite{ares2023lack,hara2026,qi2025}).
It is therefore natural to ask whether one can assess the symmetry properties of the state without leaving the Gaussian manifold. Such an approach would allow one to study symmetry restoration by measuring the distance to a set of states that are, in principle, dynamically connected to $\rho$.
Beyond its conceptual appeal, this approach would also be of significant analytical value, as it would allow the study of asymmetry using only correlation-matrix techniques~\cite{Peschel_2003}. As is well known, within the standard formulation of EA these methods only provide access to the charged moments~\cite{ares2023entanglement}. In contrast, remaining within the Gaussian manifold would open the way to closed-form analytical expressions for the large space–time dynamics of the asymmetry, based on the quasiparticle picture~\cite{quench2,alba1,alba2,calabrese2020}, which is known to emerge in a broad class of free quenches from Gaussian states \cite{fagotti2008evolution,Bertini_2018,alba2019entanglement,alba2019quantum,alba_carollo2021,parez2021quasiparticle,Parez_2022,entanglement_transition_quasiparticles,rottoli2024entanglementhamiltoniansquasiparticlepicture,travaglino2025,travaglino2026dissipative}.
In this work we face precisely this problem, and introduce a measure of asymmetry which lies within the Gaussian manifold. 
We then show that this framework can be used as a diagnostic for both the Mpemba effect and the absence of symmetry restoration. In both cases, it yields simple, asymptotically exact results within the quasiparticle picture, allowing for an immediate criterion for the occurrence of the Mpemba effect that shares the same semplicity as the one in Ref. \cite{ares2025simpler}.
Finally, we comment on the idea to use charge fluctuations as a probe of symmetry breaking.

\section{Some notions on Gaussian states}
\label{sec:gaussian}
We first briefly introduce the main ideas regarding Gaussian states which will be used in the following. A (fermionic) Gaussian state is the ground, excited, or thermal state of a quadratic hamiltonian. 
Equivalently, it is a state in which all multi-point correlation functions can be reduced to the two-point functions, which can be encoded in the correlation matrix 
\begin{equation} 
C = \begin{pmatrix}
    G && F\\F^\dag && 1 -G^T
\end{pmatrix}
\end{equation}
where $G = \braket{c_i^\dag c_j}_{i,j=1,...L}$ and $F = \braket{c_i c_j}_{i,j=1,...L}$ represent the normal and pairing correlators respectively, $c_i$ and $c^\dag_i$ being lattice spinless fermions defined by the canonical anticommutation relations $\{c_i,c_j^\dag\}=\delta_{i,j}$ (the approach is straightforwardly generalised to continuous fermions \cite{Calabrese_2011}, but we focus here on lattices for simplicity).  
Since a Gaussian state is fully determined by the correlation matrix $C$, we will refer to it as $\rho_C$. Reduced density matrices for any subsystem $A$ are also Gaussian, and obtained from the restriction $C_A$ of the correlation matrix to the subsystem. When restricting to the subsystems, the relation between the state and the correlation matrix can be understood in terms of the Entanglement Hamiltonian (EH) defined as $\rho_A = e^{K_A}/Z_A$ \cite{ehFF1,EHFF2,EH,Cardy_2016}, which can be expressed through the relation \cite{PhysRevB.69.075111,Peschel_2009}
\begin{equation}
    (k_A)_{ij} = \log(1-C_A^{-1})_{ij}
\end{equation}
where $(k_A)_{ij}$ are the matrix elements of the EH in the natural fermionic basis.
 
It is easy to show that for a Gaussian state described by a given correlation matrix $C_A$, $[\rho_A,Q_A] = 0 \iff F_A=0$. The $\Rightarrow $ implication is immediately obtained by considering the properties of $\Tr[\rho_A c_i c_j]$ under rotations of $\rho$ induced by the group elements: by the symmetry, we have 
\begin{equation}
 \Tr[\rho_A c_i c_j] = \Tr[e^{i\alpha Q_A} \rho_A e^{-i\alpha Q_A} c_i c_j ] =    e^{2i\alpha} \Tr[\rho_A c_i c_j]  
\end{equation}
where we have used cyclicity of the trace and $e^{-i\alpha Q_A} c_i e^{i\alpha Q_A} = e^{i\alpha }c_i$; the equivalence can thus only be satisfied if $F_A=0$ identically.
The opposite implication comes instead from the consideration that if $F_A=0$ the Entanglement Hamiltonian is of the form   $ K_A=\sum_{x,y \in A} k_{xy} c_x^\dag c_y $ \cite{Peschel_2009} and therefore commutes with $Q_A$, which immediately implies that $\rho_A$ commutes with $Q_A$. 
It is important to note that the condition $[\rho_A, Q_A] = 0$ represents a weak form of symmetry, as it does not imply that $\rho_A$ is an eigenstate of the charge operator. Indeed, the reduced density matrix is typically mixed, and for this reason even a (weakly) symmetric state of this kind can exhibit nontrivial charge fluctuations within the subsystem $A$.

Given a non-Gaussian state $\rho_A$, it is always possible to Gaussianise it: this is simply done by constructing the (unique) Gaussian state $\rho_{C_A}$ from the correlation matrix $C_A$ of $\rho_A$. This allows to introduce a natural measure of Non-Gaussianity \cite{genoni2008,genoni2010,Ares:2026jes}
\begin{equation}
    {\rm NG}(\rho) = S(\rho_A||\rho_{C_A}) = S(\rho_{C_A}) - S(\rho_A).
\end{equation}
This is a good measure, as it minimises the distance between $\rho_A$ and the manifold of Gaussian states \cite{marian2013}. It is possible to highlight an interesting relation between non-Gaussianity and asymmetry:
given a Gaussian state $\rho_{C_A}$ defined by a correlation matrix
\begin{equation} 
\label{eq:CA}
C_A = \begin{pmatrix}
    G_A && F_A\\F_A^\dag && 1 -G_A^T
\end{pmatrix},
\end{equation}
its symmetrisation $ \sum_q P_q \rho_{C_A} P_q$ is a strongly non-Gaussian state. However the symmetrisation acts on the correlation matrix in a simple way,
\begin{equation}
    \begin{split}
        \Tr\left[\left( \sum_q P_q \rho_{C_A} P_q \right)c_i^\dag c_j\right] &=  \Tr\left[  \rho_{C_A} \left( \sum_q P_q c_i^\dag c_j P_q\right)\right] =  \Tr[  \rho_{C_A}  c_i^\dag c_j ] \\
         \Tr\left[ \left(\sum_q P_q \rho_{C_A} P_q\right) c_i c_j\right] &=  \Tr\left[  \rho_{C_A} \left( \sum_q P_q c_i c_j P_q\right)\right] = 0, 
    \end{split}
\end{equation}
namely, the normal part is unvaried while the pairing part is set to zero. Therefore, the Gaussianisation of the symmetrised state is simply a Gaussian state $ \rho_{C_A^{(s)}}$ identified by the correlation matrix
\begin{equation}
    C_A^{(s)} = \begin{pmatrix}
    G_A && 0\\0 && 1 -G_A^T
\end{pmatrix},
\end{equation}
which, as shown above, is a symmetric state.
It is therefore natural to introduce the notion of \emph{Gaussian symmetrisation}, defined as an operation that maps a generic Gaussian state $\rho_{C_A}$ to a Gaussian state $\rho_{C_A^{(s)}}$ by directly symmetrising its correlation matrix. We stress that this Gaussian symmetrisation is always well defined, since the only conditions required for $G_A$ and $F_A$ to define a valid Gaussian state are that $G_A = G_A^\dagger$ and $F_A = -F_A^T$, as dictated by the fermionic structure, that $G_A$ has eigenvalues in $[0,1]$, and that $G_A(1 - G_A) \geq F_A F_A^\dagger$, ensuring that the full correlation matrix $C_A$ has eigenvalues in $[0,1]$. 
Therefore, it is always possible to remove the pairing term $F_A$ without violating these constraints. 
By contrast, the opposite construction, namely defining a state with only pairing correlations (which would correspond to a state that completely breaks the symmetry, in the sense that its symmetrisation vanishes), is not possible. 
Our construction of Gaussian symmetrisation highlights an interesting structure:  defining the three transformations as $\mathcal{S}:\rho\to \rho_Q$,  $\mathcal{G}:\rho\to \rho_C$ and $\mathcal{S}^{(G)} : \rho_C \to \rho_{C^{(s)}}$, it is clear that $\mathcal{S}$ and $\mathcal{G}$ do not commute. However we have that
\begin{equation}
    \mathcal{G(S(\rho))} = \mathcal{S}^{(G)} (\mathcal{G}(\rho))  =\mathcal{G}( \mathcal{S} (\mathcal{G}(\rho)) ).
    \label{comp}
\end{equation}
In the following sections we will show that this equation provides the natural way to define a fully Gaussian asymmetry measure.

\section{Entanglement asymmetry in the Gaussian manifold}
 For simplicity of notation, since all states we consider from now on are Gaussian, we drop the nomenclature of the previous section, and we refer to a generic Gaussian state in $A$ as $\rho_A$ and to a symmetric state as $\rho_A^{(s)}$. 
 As highlighted in the above, given a state $\rho_A$ described by a correlation matrix \eqref{eq:CA} it is natural to consider a definition of symmetrisation at the level of the correlation matrix itself; this leads to a symmetrised state defined by the correlation matrix
\begin{equation}
    C_A^{(s)} = \begin{pmatrix}
    G_A && 0\\0 && 1 -G_A^T
\end{pmatrix}.
\end{equation}
A natural quantifier of the ``distance" between the two states is given by the relative entropy, as in the general definition of asymmetry \cite{ares2023entanglement},
\begin{equation}
    \Delta S^{(G)}_A = S(\rho_A||\rho_A^{(s)}) = \Tr[\rho_A (\log\rho_A - \log \rho_A^{(s)})].
    \label{eq:gaussianasymmetry}
\end{equation}
where the label $(G)$ stresses that this is a Gaussian definition of the entanglement asymmetry. 

One of the main features of the standard EA is that it reduces to the difference of the entropies evaluated in the two states: we can show that the same is valid also in the present case, although this originates from a different mechanism. As discussed previously, the EH corresponding to $\rho_A^{(s)}$  is quadratic and without pairing terms, $\log\rho_A^{(s)} \sim \sum_{x,y} A_{x,y} c_x^\dag c_y$, hence
\begin{equation}
    \Tr[\rho_A \log \rho_A^{(s)}] \sim \sum_{x,y} A_{x,y} \braket{c_x^\dag c_y}_{\rho_A}.
\end{equation} 
Since the two states $\rho_A$ and $\rho_A^{(s)}$ share the normal component of the correlation matrix $G_A$, the two-point function is the same when evaluated over the two different density matrices, $ \braket{c_x^\dag c_y}_{\rho_A}= \braket{c_x^\dag c_y}_{\rho_A^{(s)}} $, implying immediately
\begin{equation}
     \Tr[\rho_A \log \rho_A^{(s)}]=\Tr[\rho_A^{(s)} \log \rho_A^{(s)}].
\end{equation}
Hence we conclude that the relative entropy simplifies to 
\begin{equation}
     \Delta S^{(G)}_A = S(\rho_A||\rho_A^{(s)}) =   S(\rho_A^{(s)}) - S(\rho_A). \label{eq:separationofentropies}
\end{equation}
Note that this result is exact, in contrast, for example, to the probe introduced in Ref.~\cite{ares2025simpler}, for which this property holds only approximately; we will however show that the two measures share some common features in the quench evolution. Since \eqref{eq:separationofentropies} only involves entropies of Gaussian states, it can be immediately evaluated through correlation matrix techniques \cite{Peschel_2009},
\begin{equation}
\label{eq:entropyfromcorrelationmatrix}
    S(C_A) = -\frac{1}{2}\operatorname{Tr}[C_A \log C_A + (1-C_A) \log (1-C_A )],
\end{equation}
which avoids completely the necessity of complicated integrals of charged moments as in the standard EA framework \cite{ares2023entanglement}.

It is also possible to prove that $\rho_A^{(s)}$ minimises the distance of $\rho_A$ from the manifold of symmetric Gaussian states, justifying its introduction as a good asymmetry measure. Consider a state $\sigma_A $ which is symmetric and Gaussian, and is hence of the form $\sigma_A = \frac{1}{Z_{\sigma}}e^{-\sum_{x,y} A^{\sigma}_{xy} c_x^\dag c_y}$. We have
\begin{eqnarray}
     S(\rho_A||\sigma_A) &=& \Tr[\rho_A (\log\rho_A - \log\sigma_A)]\nonumber \\
     &=&\Delta S^{(G)}_A + \Tr[\rho_A (\log\rho_A^{(s)}-\log\sigma_A )]
\end{eqnarray}
Since both $\rho_A^{(s)}$ and $\sigma_A$ are symmetric Gaussian states, the expectation value of their logarithm is of the form 
\begin{equation}
     \Tr[\rho_A (\log\rho_A^{(s)}-\log\sigma_A )] \sim \sum_{x,y}\left(A^{\rho}_{xy} -A_{xy}^{\sigma} \right) \braket{c_x^\dag c_y}_{\rho_A}.
\end{equation} 
Using again the fact that $\rho_A$ and $\rho_A^{(s)}$ share the same normal part of the correlation matrix, we can apply the same simplification
\begin{equation}
    \sum_{x,y}A_{xy}^{\rho/\sigma} \braket{c_x^\dag c_y}_{\rho_A} = \sum_{x,y}A_{xy}^{\rho/\sigma} \braket{c_x^\dag c_y}_{\rho_A^{(s)}}
\end{equation}
allowing us to rewrite 
\begin{equation}
     \Tr[\rho_A (\log\rho_A^{(s)}-\log\sigma_A )] = \Tr[\rho_A^{(s)} (\log\rho_A^{(s)}-\log\sigma_A )].
\end{equation}
This is the relative entropy between the two symmetric states, which is non-negative and zero only if $\sigma_A = \rho_A^{(s)}$, thus completing the proof. Hence, we conclude that $\Delta S^{(G)}_A$ minimises the relative entropy of any Gaussian state with the manifold of symmetric Gaussian states.
We stress that in general $\Delta S^{(G)}_A$ is much larger than the standard asymmetry $\Delta S_A$. 
In fact, while it is proven that $\Delta S_A \leq \log \ell_A$ \cite{Ares:2026jes}, we will show in the following that the Gaussian asymmetry can be (and in general is) extensive in $\ell_A$.  This property implies that the standard symmetrised state $\rho_Q$ is in general very far from a Gaussian state.

Quench dynamics from generic states typically leads to symmetry restoration of the density matrix at long times. In standard asymmetry treatments, the symmetry restoration is understood as the statement $\rho_{A} \overset{t\to \infty}{\to}  \rho_{A,Q}$.
In the present context, since the evolution of $\rho_A$ takes place within the Gaussian manifold, this restoration is perhaps more properly understood as 
\begin{equation}
    \rho_A \overset{t\to \infty}{\to} \rho_A^{(s)}
\end{equation}
Hence, the time evolution of $\Delta S^{(G)}_A$ provides an alternative route to study dynamical symmetry restoration and its main features, including the Mpemba effect  (which will be investigated in the next section). Note that there are also cases in which the state does not symmetrise, such as quenches from the tilted Néel state~\cite{ares2023lack}; this effect can likewise be captured through the behavior of Eq.~\eqref{eq:gaussianasymmetry}. 
Moreover, using $\rho_A^{(s)}$ instead of the standard symmetrisation enables the introduction of an alternative asymmetry quantifier based on the full counting statistics (FCS) of the charge, which we discuss in Sec.~\ref{sec:FCSasym}.

As a final remark, we note that, beyond the significant analytical simplifications it offers, the Gaussian asymmetry admits a natural interpretation within the framework of quantum resource theory (QRT)~\cite{gour2015,chitambar2019}. While the standard EA defines a monotone under general symmetric operations~\cite{Vaccaro2008,bartlett2007,Gour2009}, its Gaussian counterpart defines a monotone under Gaussian symmetric operations. It therefore represents the appropriate measure when the dynamics is restricted to Gaussian operations, such as evolution under quadratic Hamiltonians, linear dissipative processes, or certain measurement protocols, including the quantum trajectories associated with continuous measurements of local fermionic densities.

\section{Gaussian Mpemba effect}
\label{sec:mpemba}
In this section, we show that the Gaussian asymmetry $\Delta S^{(G)}_A$ in Eq.~\eqref{eq:separationofentropies} can be used to detect the Mpemba effect in the typical quench protocols considered in the literature. In particular, we demonstrate that $\Delta S^{(G)}_A$ admits a simple description in terms of a quasiparticle picture, valid in the regime where both $t$ and $\ell_A$ are large compared to microscopic scales, yet remain much smaller than the total system sise.
We focus on the dynamics governed by hopping model $H= -\frac{1}{2}\sum_i c_i^\dag c_{i+1} + h.c.$, which corresponds to the fermionic formulation of the XX spin-chain, and consider as initial states a general class of coherent states frequently appearing in quench dynamics \cite{rottoli2024entanglementhamiltoniansquasiparticlepicture,travaglino2024}, i.e. \begin{equation}
\label{eq:coherent}
    \ket{\psi}\propto \exp\left\{\sum_k M(k) c_k^\dag c_{-k}^\dag\right\}\ket{0}
\end{equation}
These states exhibit a pair structure by construction and therefore provide the simplest setting for applying the quasiparticle picture.
Since in such quenches the normal part of the correlation matrix is time-independent, 
\begin{equation}
    \braket{c_k^\dag(t) c_q(t)} = \braket{c_k^\dag c_q } =\delta_{k,q} \frac{|M(k)|^2}{1+|M(k)|^2},
\label{eq:occ_coherent}
\end{equation}
we immediately see that $G_A$ does not evolve in time, and therefore $\rho_A^{(s)}$ remains constant. 
On the other hand, the time-dependent phases in the pairing terms
\begin{equation}
    \braket{c_k^\dag(t) c^\dag_q(t)}  =\delta_{q,-k} e^{-2it\varepsilon_k} \frac{M(k) }{1+|M(k)|^2},
\end{equation}
lead to cancellations in the real-space correlators $\braket{c_x^\dag c_y^\dag}$ via stationary phase cancellations, implying that $F_A(t) \to 0$ as $t\to \infty$. 
Consequently, $\rho_A^{(s)}$ is  identical to $\lim_{t \to \infty}\rho_A(t)$ and 
    \begin{equation}
  S(\rho_A^{(s)})=S(\rho_A(\infty)).
    \end{equation}
Hence $S(\rho_A^{(s)})$ corresponds to the volume law entropy arising in the long time limit in the quench dynamics of $\rho_A$.
Using that $S(\rho_A)$ admits the quasiparticle formulation \cite{quench2,alba1,alba2,rottoli2024entanglementhamiltoniansquasiparticlepicture}
\begin{equation}
    S(\rho_A(t))= \int \frac{dk}{2\pi} \min(2|v_k|t,\ell_A) s_k(n_k),
    \label{eq:standard_qpp}
\end{equation}
 where $s_k = -\left(n_k\log n_k + (1-n_k) \log(1-n_k))\right) $ is the pair contribution to the entanglement entropy,
the Gaussian asymmetry is immediately obtained as
\begin{equation}
     \Delta S^{(G)}_A = \int \frac{dk}{2\pi} \max(\ell_A - 2|v_k|t,0)  s_k,
     \label{eq:qpp_mpemba}
  \end{equation}
Eq.~\eqref{eq:qpp_mpemba} provides a criterion for the onset of the Mpemba effect. 
Indeed, since $s_k$ is maximal when the occupation function is close to $1/2$ and vanishes when it approaches $0$ or $1$, and since the factor $\max(\ell_A - 2|v_k|t, 0)$ suppresses the contribution of fast modes first, it suffices to examine the relative purity of slow and fast modes. In particular, the Mpemba effect arises when the slow modes (with momenta $k \sim 0,\pi$) are “purer”, in the sense that their occupation functions are closer to $0$ or $1$, compared to the fast modes (with momenta $k \sim \pm \pi/2$). 
This reasoning is in the same spirit as the discussion of Ref. \cite{rylands2024microscopic}, in which the origin of the Mpemba effect was related to the amount of charge carried by fast and slow modes. 
In the present case, the charge carried by each excitation is precisely given by the occupation function, which uniquely determines the entanglement. This is a key feature of the quasiparticle picture: since entanglement is transported by particle-like excitations, its content is directly encoded in the occupation function. As discussed in Refs.~\cite{travaglino2024,Travaglino2025measurements}, the occupation function can be interpreted as the single-particle charge eigenvalue associated with coarse-grained quasiparticle excitations, which are responsible for the propagation of entanglement.

\begin{figure}[t]
    \centering
    \includegraphics[width=\linewidth]{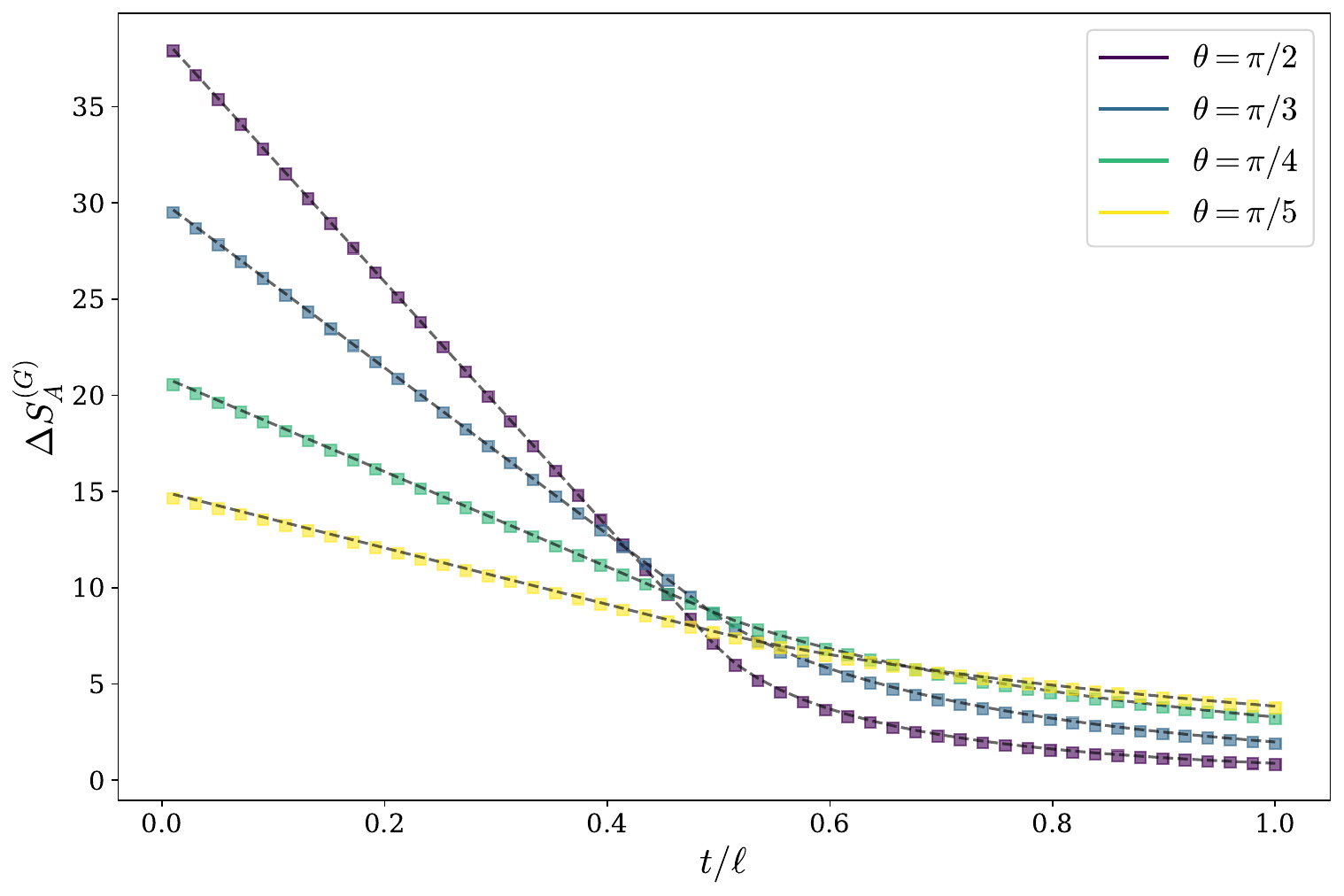}
    \caption{Gaussian asymmetry in a quench from tilted ferromagnetic state for a subsystem of size $\ell$ with different values of tilting angle $\theta$. 
    The curves clearly show the occurrence of the quantum Mpemba effect. 
    Symbols are numerical exact results at $\ell_A=100$ and lines are the quasiparticle picture prediction. 
    The linear decrease up to $t=\ell_A/2$ and the following saturation is a clear sign of the validity of the quasiparticle picture directly at the level of $\Delta S^{(G)}_A$. This has to be contrasted with the standard EA, which is nonlinear at early times.}
    \label{fig:tiltedferro}
\end{figure}

We now explore the occurrence of the Mpemba effect in the dynamics of the Gaussian asymmetry $\Delta S^{(G)}_A$ 
in a quench from a tilted ferromagnetic state, which is of the form \eqref{eq:coherent}.
This specific quench is characterised by velocities $v_k= \varepsilon'_{k}$ and occupation functions \cite{murciano2024entanglement}
\begin{equation}
    n_k^{(\theta)} = \frac{1}{2}\left(1-\frac{2\cos\theta-(1+\cos^2\theta) \cos k}{1+ \cos^2\theta -2 \cos\theta \cos k}\right).
    \label{eq:occtiltedferro}
\end{equation}
The results for the Gaussian asymmetry are reported in Fig. \ref{fig:tiltedferro}: the symbols represent the exact solution obtained through \eqref{eq:entropyfromcorrelationmatrix}, while the dashed lines are the quasiparticle prediction \eqref{eq:qpp_mpemba}, showing excellent agreement even for relatively small subsystem size $\ell_A$.
While the qualitative picture is analogous to that of the standard asymmetry discussed in Ref.~\cite{ares2023entanglement}, the behavior of $\Delta S^{(G)}_A$ differs in important ways. In particular, its initial value is extensive in $\ell_A$, and the linear decrease observed at early times provides a clear signature of the quasiparticle picture. This feature is notably absent in the case of the standard EA.

We note that, in the present case, due to the absence of time dependence in $\rho_A^{(s)}$, the Gaussian asymmetry $\Delta S^{(G)}_A$ reduces to the quantity considered in Ref.~\cite{ares2025simpler}, where a probe of the Mpemba effect is defined as the relative entropy between the time-evolving state and the asymptotic state $\rho_A(\infty)$. 
It is important to emphasise, however, that while $\Delta S^{(G)}_A$ is a bona fide asymmetry measure in the sense discussed above, the quantity introduced in Ref.~\cite{ares2025simpler} should rather be viewed as an efficient probe of the Mpemba effect, and does not necessarily share the general properties of an asymmetry measure. 
As we will show in the next subsection, this equivalence is specific to the class of states considered here.

\subsection{Lack of symmetry restoration}
While for states of the form \eqref{eq:coherent} considered in the previous subsection the Gaussian asymmetry $\Delta S^{(G)}_A$ reduces to the relative entropy with respect to the asymptotic state, it is possible to consider a much broader class of initial states. In particular, one can study states with a pair structure expressed in terms of Bogoliubov fermions $\eta_k$, rather than the original fermionic operators. Since the charge $Q_A$ counts the number of the original fermions, this mismatch leads to a particularly rich and nontrivial phenomenology.
A representative example is provided by the tilted Néel state, studied in Ref.~\cite{ares2023lack}, where it was shown that such an initial condition leads to a lack of symmetry restoration. This can be understood from the fact that, at long times, the state does not relax to $\sum_q P_q \rho_A P_q$, or equivalently to $\rho_A^{(s)}$.
In this case, constructing a quasiparticle description for the Gaussian asymmetry $\Delta S^{(G)}_A$ is more involved, as the symmetrised state no longer possesses a time-independent correlation matrix. Nevertheless, the entropy $S(\rho^{(s)})$ can still be described within a quasiparticle framework, consistent with the Gaussian nature of the state, in which quasiparticles propagate on top of a background carrying a finite entropy density. 
The construction is analogous to the one of Ref. \cite{ares2023lack} for the charged moments, and involves a quasiparticle picture ansatz for the difference between the entropy at time $t$ and that at time 0, i.e.
\begin{equation}
     S(\rho_A^{(s)}(t))- S(\rho_A^{(s)}(0)) = \int \frac{dk}{2\pi} \min(2|v_k|t,\ell_A) (s_k(n^{(\infty)}_k) - s_k (n^{(0)}_k))
     \label{eq:qpponbackground}
\end{equation}
where the population $n_k^{(\infty)}$ and  $n_k^{(0)}$ fix the volume law entanglement at times zero and infinity, in the sense that 
 \begin{equation}
     S(\rho_A^{(s)}(0 )) = \ell_A \int \frac{dk}{2\pi}  s_k (n^{(0)}_k), \hspace{0.5cm} S(\rho_A^{(s)}(\infty )) = \ell_A \int \frac{dk}{2\pi}  s_k (n^{(\infty)}_k).
 \end{equation}
These population can be determined case by case, as it will be done below for the specific choice of the tilted Néel state.
Eq. \eqref{eq:qpponbackground} can be rewritten as  
 \begin{equation}
 \begin{split}
     S_A(\rho_A^{(s)}(t)) = \int \frac{dk}{2\pi} \min(2|v_k|t,\ell_A) s_k(n_k^{(\infty)}) + \int \frac{dk}{2\pi} \max(\ell_A-2|v_k|t,0 )s_k(n_k^{(0)})
 \end{split}
 \end{equation}
where the two contributions admit an interpretation in terms of shared pairs (in which one quasiparticle is in $A$ and the other is in the complement) and pairs which are fully in $A$, see also the discussion of Refs. \cite{Travaglino2025measurements,travaglino2026dissipative}.
Clearly, this expression reduces to the same result as in the previous subsection if $S(C_A^{(s)}(t))= S(C_A^{(s)}(0))$, in which case the correlation matrix is constant and therefore fixed by the final state. On the other hand, the solution for the standard entropy $S(\rho_A)$ simply admits the standard quasiparticle picture \eqref{eq:standard_qpp}, with occupation functions $n_k$. This reasoning leads to the final result 
\begin{equation}
        \Delta S^{(G)}_A  = \int \frac{dk}{2\pi} \min(2|v_k|t,\ell_A) \left( s_k(n_k^{(\infty)}) - s_k(n_k)\right)+ \int \frac{dk}{2\pi} \max(\ell_A-2|v_k|t,0 )s_k(n_k^{(0)}).
        \label{eq:qpptiltedneel}
\end{equation}
The latter equation can be specialised to the tilted Néel state using the techniques developed in Ref.~\cite{ares2023lack}. 
In terms of the functions 
\begin{equation}\begin{split}
    g_{11}(k) &= -\cos\theta - \frac{\cos\theta \sin^2 \theta  (\cos k +\cos^2\theta )}{1+2\cos k \cos^2\theta + \cos^4\theta},\\
    g_{12}(k) &= - \frac{\cos(k/2) (1-\cos^4\theta)}{1+2\cos k \cos^2\theta + \cos^4\theta},\\
    f_{12}(k) &= -\frac{\cos \theta \sin^2\theta \cos k}{1+2\cos k \cos^2\theta + \cos^4\theta}.
    \end{split}
\end{equation}
the three occupations are $n^{(\infty)}_k= (1+g_{12})/2$, $n^{(0)}_k = (1+\sqrt{g_{11}^2 + g_{12}^2})/2$, and finally $n_k = (1+ g_{12} + f_{12})/2$. Moreover, the velocity of the quasiparticles is $v_k = \varepsilon'_{k/2}$ because of the two-site shift invariance \cite{ares2023lack}.

\begin{figure}[t]
    \centering
    \includegraphics[width=\linewidth]{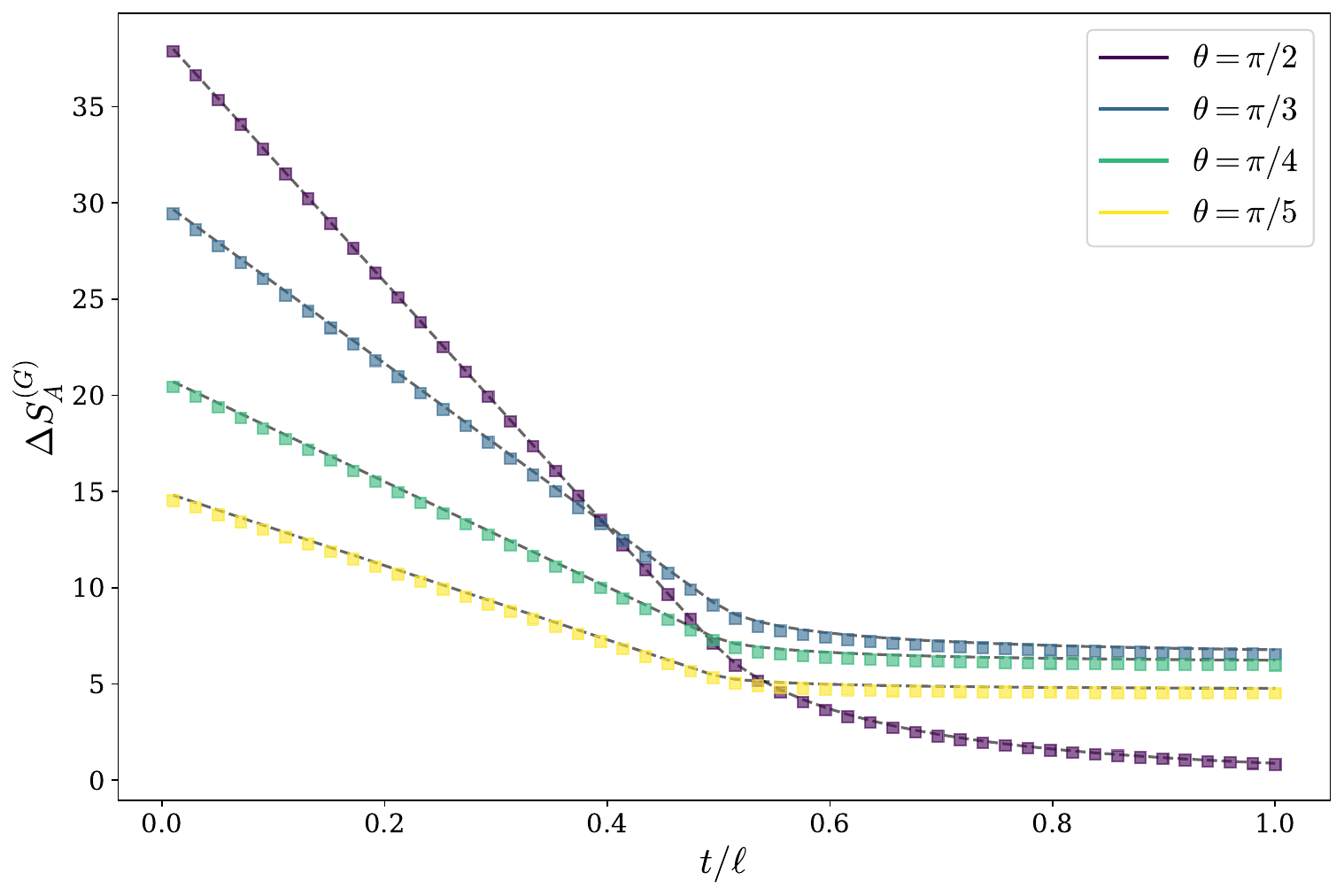}
    \caption{Gaussian asymmetry as a probe for the lack of symmetry restoration in a quench from the tilted Néel state with different tilting angles. 
    The asymmetry is evaluated over a subsystem of size $\ell=100$. As in Fig. \ref{fig:tiltedferro} notice the linear decrease followed by saturation. }
    \label{fig:tiltedneel}
\end{figure}

The resulting dynamics is shown in Fig.~\ref{fig:tiltedneel}, which demonstrates the excellent agreement between the exact prediction obtained from Eq.~\eqref{eq:entropyfromcorrelationmatrix} and the quasiparticle result in Eq.~\eqref{eq:qpptiltedneel}. 
In particular, it is evident that, at long times, the system generally does not restore the symmetry. The origin of this behavior is clearly captured by the quasiparticle expression~\eqref{eq:qpptiltedneel}. Indeed, in the long-time limit the second line vanishes due to the structure of the counting function, while the first term remains generically non-zero. From the explicit form of $f_{12}(k)$, it follows that this contribution vanishes only at the special points $\theta = 0, \frac{\pi}{2}$, where symmetry restoration is indeed recovered.
This behavior admits a simple physical interpretation. As discussed in Ref.~\cite{ares2023lack}, the quasiparticles forming the pair structure are Bogoliubov fermions $\eta_k$, obtained from a mixing of $c_k$ and $c_k^\dagger$. As a result, the occupation functions of the modes $\eta_k$ depend on both the normal component $G_A$ and the pairing component $F_A$ of the correlation matrix expressed in terms of the original fermions. Consequently, removing the $F_A$ contribution significantly alters the occupation functions of the propagating modes, leading to a mismatch between $n_k$, computed from the full correlation matrix, and $n_k^{(0)}$ and $n_k^{(\infty)}$, obtained from the truncated one. 
It is worth noting that the result of the previous subsection, Eq.~\eqref{eq:qpp_mpemba}, is recovered only when the three occupation functions coincide, which occurs precisely in the case $\eta_k = c_k$.

\section{Typical Gaussian asymmetry}
\label{sec:typical}

Beyond the quantum Mpemba physics, key results obtained with entanglement asymmetry concern its characterisation in the context of random states~\cite{ares2024blackhole,Russotto2025,Russotto_2025b,Fujimura2026,Joshi2026}, also in connection with black hole physics. 
A remarkable result is that the typical asymmetry of Haar random states exhibits a sharp discontinuity when the subsystem size reaches half of the total system size.
It is then natural to investigate the typical properties of the Gaussian asymmetry $\Delta S^{(G)}_A$ when averaging over the ensemble of Gaussian states. 
We are therefore interested in 
\begin{equation}
    \braket{\Delta S^{(G)}_A}_g = \braket{S(\rho_A^{(s)})}_g - \braket{S(\rho_A)}_g
\end{equation}
where $ \braket{.}_g$ represents the average over Gaussian states, see for instance 
\cite{bianchi2021}.
This average can be accessed by following the approaches developed in Refs.~\cite{vidmar2020,bianchi2021,bianchi2022,Murciano_2022bb}, since both $\rho_A$ and $\rho_A^{(s)}$ are Gaussian. 

\begin{figure}[t]
    \centering
    \includegraphics[width=\linewidth]{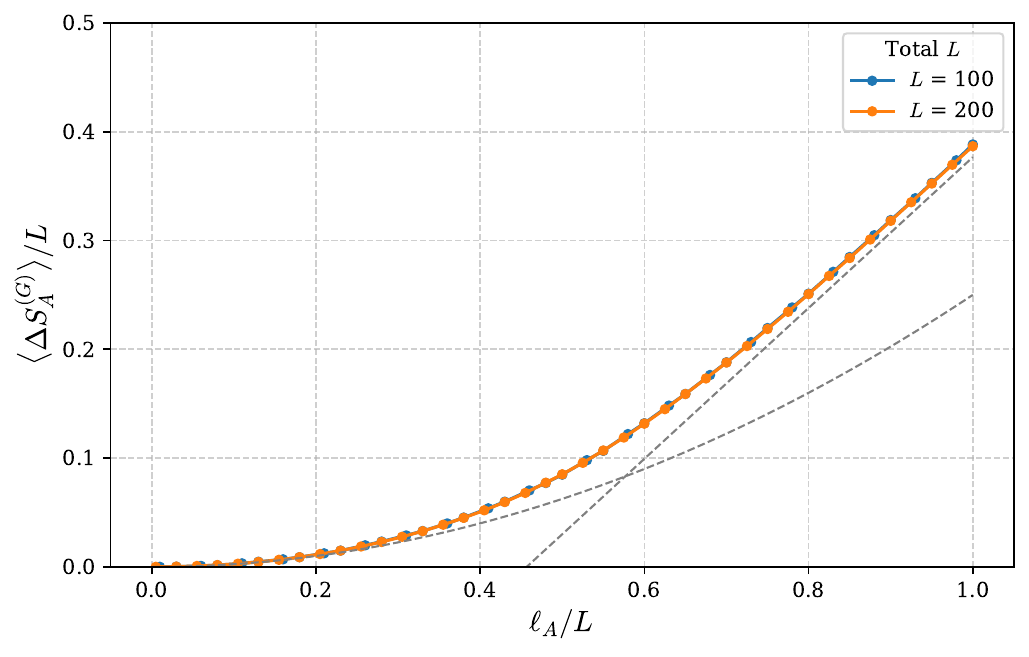}
    \caption{$\braket{\Delta S^{(G)}_A}_g$ for typical Gaussian states. The symbols represents the numerical averages over large samples for $L=100$ and $L=200$.
    The behavior is smooth as a function of $\ell_A/L$. 
    The dashed lines represent the two asymptotic regimes, confirming the parabolic growth at small $\ell_A$ and the linear behavior for $\ell_A \to L$.}
    \label{fig:random}
\end{figure}

Fig.~\ref{fig:random} shows the numerical results obtained for systems of total size $L=100,200$, averaged over a large ensemble of Gaussian states. 
The overlapping curves for both system sizes show that finite-size effects are negligible. 
The most striking feature is that the Gaussian asymmetry exhibits a smooth dependence on the subsystem size, with no indication of discontinuities, in clear contrast to the behavior observed for Haar-random states \cite{ares2024blackhole}. 
It would be interesting to investigate whether the standard entanglement asymmetry for Gaussian random states also displays such smooth behavior; however, this lies beyond the scope of the present work.

One interesting question is whether the curve displayed in Fig.~\ref{fig:random} can be obtained analytically; however, this turns out to be rather cumbersome. 
In this context, we have been able to derive the leading-order behavior in the two relevant regimes $L_A \ll L$ and $L_A \to L$.
In the regime $L_A \ll L$, it is useful to consider the expansion for the entanglement entropy of Gaussian states \cite{Peschel_2003,Peschel_2009}
\begin{equation}
    S_A =\ell_A \log 2 - \sum_{n=1}^{\infty} \frac{\Tr[C_A^{2n}]}{4n(2n-1)}
    \label{eq:entropyexpansion}
\end{equation}
In fact, for $L_A\ll L$ the leading order terms in the sum are the ones corresponding to smaller values of $n$; the first two terms give \cite{bianchi2022}  
\begin{equation}
     \braket{S(\rho)}_g \approx\ell_A \log 2 -  \frac{\ell_A^2}{2L}  +\dots
     \label{eq:averagesmallLA}
\end{equation}
In order to obtain $ \braket{S(\rho^{(s)}})$, it is useful to understand how the second term in \eqref{eq:averagesmallLA} arises from \eqref{eq:entropyexpansion}. The variance of the correlation matrix is 
\begin{eqnarray}
    \Tr[C_A^2] &=& 2\Tr[|F_A|^2] + \Tr[G_A^2] + \Tr[(1-G_A^T)^2] \nonumber\\
    &=& 2\Tr[|F_A|^2] + 2\Tr[G_A^2] + \Tr[1-2G_A]. \hspace{0.5cm}
\end{eqnarray}
Averaging over Gaussian states $\langle\Tr[1-2G_A]\rangle_g=0$, i.e. the average configuration is at half-filling. 
Hence
\begin{equation}
    \braket{\Tr[C_A^2] }_g=2 \braket{\Tr[|F_A|^2] + \Tr[G_A^2]}_g.
\end{equation}
The contributions from the two parts are equal due to the rotational invariance assumed in the construction of Ref.~\cite{bianchi2022}. 
Hence, from \eqref{eq:averagesmallLA} we conclude that
\begin{equation}
    \braket{\Tr[|F_A|^2] }_g= \braket{\Tr[G_A^2]}_g \approx \frac{\ell_A^2}{L}.
\end{equation}
This allows us to obtain 
$\braket{S(\rho^{(s)})}_g$ by considering only the contribution of $ G_A$, giving
\begin{equation}
   \braket{S(\rho^{(s)})}_g \approx\ell_A \log 2 - \frac{\ell_A^2}{4L} + \dots
\end{equation}
leading finally to
\begin{equation}
    \braket{\Delta S^{(G)}_A}_g \approx \frac{\ell_A^2}{4L}.
\end{equation}
Additional terms in the expansion for small $\ell_A/L$ can be derived systematically by including higher-order cumulants in Eq.~\eqref{eq:entropyexpansion}.

In the opposite regime, for $L_A$ close to $L$, some simple observations can be made. 
Since $\rho$ is globally a pure state, its entropy vanishes as $L_A \to L$. 
By contrast, the symmetrised Gaussian state has a large entropy even in this limit: 
the removal of the pairing terms implies that the full correlation matrix is no longer a projector, even when the subsystem coincides with the entire system. 
As a consequence, the leading-order behavior of the average entropy is
$ \braket{S(\rho^{(s)})}_g =\ell_A \log 2$, while all other terms in Eq. \eqref{eq:entropyexpansion} are subleading in $\ell_A$. 
This gives  
\begin{equation}
    \braket{\Delta S^{(G)}_A}_g \approx\ell_A\log2 +O(\ell_A^0),
   \label{DSL} 
\end{equation}
which implies that $\Delta S^{(G)}_A$ is extensive for large subsystem sizes.

Eq. \eqref{DSL}  should be contrasted with the typical behavior of the EA~\cite{ares2024blackhole}, where the scaling in this regime is logarithmic in $\ell_A$. This difference can be interpreted as indicating that $\rho_Q$ induces an almost maximal breaking of Gaussianity; we will return to this point in the next section.

\section{Relation to entanglement asymmetry and non-Gaussianity}
As emphasised throughout the previous sections, given a Gaussian state $\rho_A$, its symmetrisation 
$\rho_{A,Q} = \sum_q P_q \rho_A P_q$ is, in general, a strongly non-Gaussian state. This observation 
was one of the main motivations for introducing the Gaussian asymmetry $\Delta S^{(G)}_A$, which provides 
an asymmetry quantifier that remains entirely within the Gaussian manifold. 
Importantly, as discussed above, $\rho_A^{(s)}$ can be viewed as the Gaussianisation of $\rho_{A,Q}$. 
This highlights a natural connection between the different asymmetry measures and the role of 
non-Gaussianity.
In fact, recalling the definition of EA $\Delta S = S(\rho_Q) - S(\rho)$ and the one of non-Gaussianity ${\rm NG}(\rho_{Q})=S(\rho_Q^{(s)})-S(\rho_Q)$, and using Eq. \eqref{comp}, the Gaussian asymmetry, can be rewritten as
\begin{equation}
   \Delta S^{(G)}_A(\rho_A) =   \Delta S_A(\rho_A) + {\rm NG}(\rho_{A,Q})
    \label{eq:nongaussianity}
\end{equation}
which implies 
${\rm NG}(\rho_Q) = \Delta S^{(G)}_A - \Delta S_A$,
confirming the original insight that the difference between the two measures of asymmetry originates only from the non-Gaussianity of $\rho_{A,Q}$. 
In particular, since $\Delta S_A \leq \log \ell_A$~\cite{Ares:2026jes}, the non-Gaussianity of the symmetrised state satisfies
$NG(\rho_{A,Q}) \geq \Delta S^{(G)}_A - \log \ell_A$.
Therefore, the typical extensivity of the Gaussian asymmetry discussed in the previous section implies that symmetrisation generally induces an extensive amount of non-Gaussianity.

Eq.~\eqref{eq:nongaussianity} shows immediately that $\Delta S^{(G)}_A$ is a monotone in the sense of quantum resource theory \cite{chitambar2019}, namely
$\Delta S^{(G)}_A\bigl(\Phi(\rho_A)\bigr) \leq \Delta S^{(G)}_A(\rho_A)$,
for any symmetric Gaussian CPTP (completely positive and trace preserving) map $\Phi$. 
In fact, since the map is symmetric, it follows that
$\Delta S\bigl(\Phi(\rho_A)\bigr) \leq \Delta S(\rho_A)$.
For the second contribution, the symmetry of $\Phi$ implies that it commutes with symmetrisation, while its Gaussianity ensures that it commutes with Gaussianisation. Denoting $\tilde{\rho}_A = \Phi(\rho_A)$, we obtain
\begin{equation}
{\rm NG}(\tilde{\rho}_{A,Q}) = {\rm NG}\bigl(\Phi(\rho_{A,Q})\bigr) \leq {\rm NG}(\rho_{A,Q}),
\end{equation}
where we have used that ${\rm NG}$ is itself a monotone under Gaussian maps.

We conclude commenting on the case in which the state $\rho$ itself is not Gaussian. In this case, we first consider its Gaussianisation $\rho_C$, then its Gaussian symmetrisation $\rho_C^{(s)}$.
It is still natural to define
\begin{equation}
    \Delta S^{(G)}_A (\rho) = S(\rho||\rho_C^{(s)}),
\end{equation}
that through simple manipulations is 
\begin{equation}
     \Delta S^{(G)}_A (\rho) = \Delta S^{(G)}_A(\rho_C) +{\rm NG}(\rho),
\end{equation}
which leads to  
\begin{equation}
    \Delta S_A(\rho) = {\rm NG}(\rho) - {\rm NG}(\rho_Q) +\Delta S^{(G)}_A(\rho_C).
\end{equation}
Once again, the last equation expresses the idea that the difference between the two asymmetry measures quantifies the increase in non-Gaussianity $ {\rm NG}(\rho_Q) - {\rm NG}(\rho)$ induced by the symmetrisation.

\section{FCS Asymmetry in the Gaussian manifold}
\label{sec:FCSasym}
The behavior of the full counting statistics (FCS) of the charge 
$Q_A = \sum_{x \in A} c_x^\dagger c_x$ in quenches from symmetric 
and non-symmetric Gaussian states (or integrable states in the 
interacting case) has been extensively analysed in 
Refs.~\cite{lovas2017,parez2021exact,parez2021quasiparticle,groha08fcs,bertini2023nonequilibrium,tirrito2023,bertini_asymmetric_2024,Senese_2024,horvath2024,horvath2025}, 
and has also been experimentally measured~\cite{Joshi_2025bb}. 
In particular, these works have demonstrated that the FCS exhibits 
qualitatively distinct behavior depending on whether the initial state 
is symmetric or not, clearly highlighting the strong dependence of 
charge transport and fluctuation properties on symmetry.

Since the goal of asymmetry measures is to quantify the breaking of a symmetry in a given local state, it is natural to define a measure based on the difference between the two full counting statistics: that associated with $\rho_A$ and that associated with its symmetrised counterpart $\rho_A^{(s)}$, namely
\begin{equation}
    \Delta Z_\beta = \log \operatorname{Tr}[e^{\beta Q_A} \rho_A] -  \log \operatorname{Tr}[e^{\beta Q_A} \rho_A^{(s)}], \hspace{0.3cm}\beta>0 .
\end{equation}
Note that the ordering is opposite to that used for the entropy, i.e., we subtract the symmetrised FCS from the original one. This choice reflects the intuitive expectation that charge fluctuations in a symmetric state are typically weaker than in a symmetry-breaking state, as will be made precise in the next section.
It is important to stress that such a measure would not provide any information in the standard asymmetry scenario of Ref.~\cite{ares2023entanglement}, since the FCS of the charge is identical when evaluated in $\rho_A$ and in $\sum_q P_q \rho_A P_q$. However, this is no longer the case when considering $\rho_A^{(s)}$. Indeed, this FCS-based measure captures differences between the two states at the level of the charge distribution, which is, in principle, directly accessible experimentally. In other words, the state can be distinguished from its Gaussian symmetrisation simply by analysing the statistics of the charge within the subsystem.
We emphasise, however, that this quantity is not, strictly speaking, a proper measure of distance from the manifold of symmetric Gaussian states, as $\rho_A^{(s)}$ does not generally minimise it. Rather, it should be viewed as a useful diagnostic that naturally emerges once it is recognised that $\rho_A^{(s)}$ minimises the relative entropy.

\subsection{Some basic properties}
In this section, we show that the difference between the two FCS exhibits key features of an asymmetry quantifier, namely that it is non-negative and vanishes if and only if the original state is symmetric.
The FCS is written in terms of the correlation matrix as \cite{Klich_2014,groha08fcs} (see also Appendix G of \cite{bertini_asymmetric_2024})
\begin{equation}
    \log Z_C(\beta) = \log \operatorname{Tr}[e^{\beta Q_A}] + \frac{1}{2}\log\det\left[\frac{1}{2}\Big(1_{2\ell_A}+(2C_A-1_{2\ell_A}) \Gamma_N\Big)\right]
\end{equation}
where $\Gamma_N = \tanh(\beta/2)\, \sigma_z \otimes1_{\ell_A}$.  Considering that the first term is independent on the state, to show that $\log Z_C \geq \log Z_{ C^{(s)}}$ we need to discuss the behavior of the determinants. The matrix of interest can be rewritten as
\begin{equation}
   \frac{1}{2}\Big(1_{2\ell_A}+(2C_A-1_{2\ell_A}) \Gamma_N\Big) =  \begin{pmatrix}
       A && B \\C && D
   \end{pmatrix}  
\end{equation}
where we have introduced the matrices $A=\frac{1}{2}+\frac{1}{2}\tanh(\beta/2) (2G_A-1) $, $B= - \tanh(\beta/2)F_A $, $C =\tanh(\beta/2) F_A^\dag $ and $D =\frac{1}{2}+\frac{1}{2}\tanh(\beta/2) (1-2G^T_A) $ (where all numbers are assumed to be multiplied by an identity $1_{\ell_A}$). The determinant can be simplified as
\begin{equation}
    \det\begin{pmatrix}
       A && B \\C && D
   \end{pmatrix} = \det A  \det(D - CA^{-1}B) .
\end{equation}
On the other hand, for the correlation matrix $C^{(s)}_A$ the terms $B$ and $C$ are zero, while $A$ and $D$ are identical to the ones above. Therefore the inequality we need to prove is 
\begin{equation}
    \det A  \det(D - CA^{-1}B)  \geq \det A  \det D  
\end{equation} 
and hence $\det(D - CA^{-1}B)  \geq  \det D$,  because $\det A>0$. This can be immediately proven noticing that
\begin{equation}
     -CA^{-1}B = \tanh\left(\frac{\beta}{2}\right)^2 F_A^\dag \left(\frac{1}{2}+\frac{1}{2}\tanh(\beta/2) (2G_A-1)\right) F_A
\end{equation}
where the numerical prefactor is strictly positive for $\beta\neq0$, the term in the bracket is strictly  positive definite for all $\beta \neq 0$ as its eigenvalues are in the region $[(1+e^{\beta})^{-1},(1+e^{-\beta})^{-1}]$ and therefore $ -CA^{-1}B$ is a positive semi-definite operator. Since $D-CA^{-1}B$ is the sum of a positive definite and a positive semi-definite operator, it follows that 
\begin{equation}
    \det(D-CA^{-1}B)\geq\det D. 
\end{equation}
In particular, the positive-definiteness of A implies that the equality can only be obtained when $F_A=0$ (such that $F_A^\dag A F_A =0$), namely when the state itself is symmetric. We therefore conclude that $\Delta Z_C(\beta)$ is a good asymmetry measure.

\subsection{Quench dynamics}
In this section we consider the quench dynamics of the FCS-asymmetry in the two classes of initial states considered above. For coherent states of the form \eqref{eq:coherent}, the calculation is identical to the one for the Gaussian asymmetry, leading to 
\begin{equation}
    \Delta Z_\beta(t) = \int \frac{dk}{4\pi} \max(\ell_A-2|v_k|t,0) \,z_\beta(k)
    \label{Zb}
\end{equation}
where the pair contribution is given by \cite{bertini_asymmetric_2024} 
\begin{equation}
    z_\beta(k) = \log\frac{1-n_k+n_ke^{2\beta}}{(1-n_k+n_ke^\beta)^2}.
\end{equation}
In the long-time limit, $\Delta Z_\beta(t)$ vanishes, as the factor $\max(\ell_A - 2|v_k|t,0)$ goes to zero for all modes, indicating that the symmetry is ultimately restored. 
By contrast, at finite times the structure of the solution reveals a qualitative distinction between symmetric and non-symmetric initial states. In the latter case, charge cumulants are extensive at short times, whereas for symmetric states they grow linearly in time (with the exception of the average charge), leading to the scaling $\ell_A - 2|v_k|t$ at early times. Note also that $z_\beta(k)$ is strictly positive for any choice of $n_k \in [0,1]$, confirming the general result proved above.

\begin{figure}[t]
    \centering
    \includegraphics[width=\linewidth]{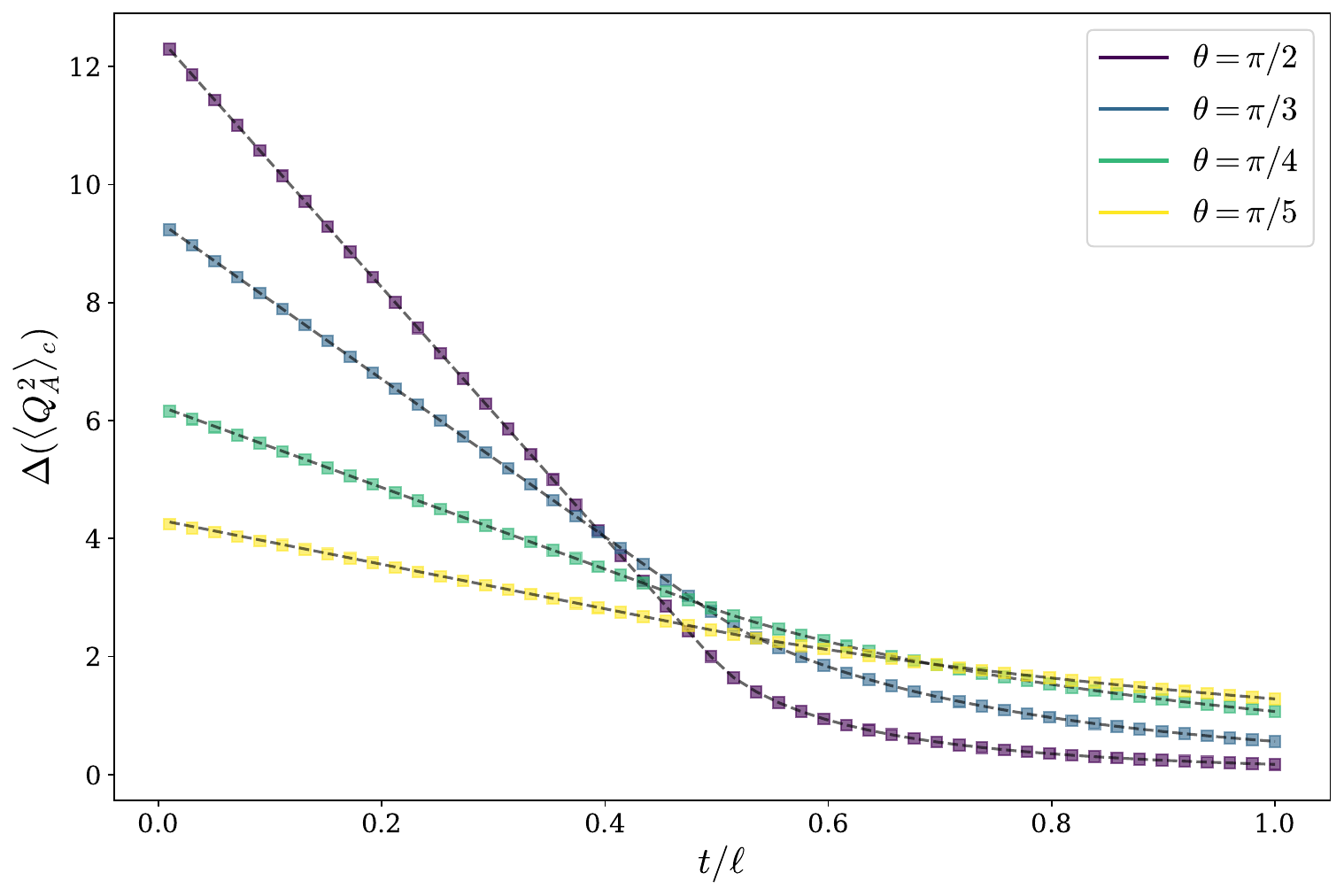}
    \caption{Dynamics of the variance difference \eqref{eq:cumulant} after a quench from the tilted ferromagnetic state. The squares are the exact predictions obtained through correlation matrix techniques, cf. Eq. \eqref{eq:deltavariances}. Dashed lines are the quasiparticle predictions \eqref{eq:cumulant}. The data show that the Mpemba effect can be diagnosed  through  the variance difference of the charge distribution.}
    \label{fig:variance_of_charge}
\end{figure}

Eq.~\eqref{Zb} can be used to evaluate the difference between the cumulants of $Q_A$ in the two states. 
The average does not provide any information because it is the same in the two states,  as it depends only on the diagonal part of the correlation matrix. Starting at the level of the variance, one obtains the nontrivial expression
\begin{equation}
    \Delta (\braket{Q^2_A}_c)(t) =  \partial_\beta^2  \Delta Z_\beta(t){\large|}_{\beta=0}=\int \frac{dk}{2\pi} \max(\ell_A-2|v_k|t,0)\, n_k (1-n_k)
    \label{eq:cumulant}
\end{equation}
which, e.g.,  can be evaluated using the tilted ferromagnet occupation functions \eqref{eq:occtiltedferro}. The difference between the two variances has similar features to the difference of the FCS, as it is $\geq 0$ and the equality is only obtained if the state is symmetric: given the general expression $ \braket{Q^2_A(C_A)}_c = \operatorname{Tr}[G_A(1-G_A) + F_A F_A^\dag]$, the difference between the two variances is the trace of a positive semi-definite operator,
\begin{equation}
    \Delta \braket{Q^2_A}_c = \operatorname{Tr}[F_A F_A^\dag]  \geq 0
    \label{eq:deltavariances}
\end{equation}
hence it is itself a good quantifier. Note that $\sqrt{\operatorname{Tr}[F_A F_A^\dag]}$ induces a semi-metric in the space of correlation matrices, which allows to define a measure of distance between the states: this is the natural candidate to introduce a measure of asymmetry which is also actually a (semi)distance, contrarily to the relative entropy which is only a divergence. It is only a semi-metric because it does not distinguish between states which have the same pairing terms, but different symmetric terms, and so it rather gives the distance between different equivalence classes, each characterised by fixed amount of symmetry breaking ($F_A$) but with generic value of symmetry preserving part ($G_A$). However, $\Delta (\braket{Q^2_A}_c) $ is not  a perfect measure of the asymmetry properties, since these are not fully encoded in $F_A$ alone (as can be, e.g., inferred  from Eq. \eqref{eq:gaussianasymmetry}).

Returning to the quasiparticle prediction~\eqref{eq:cumulant}, we note that the function $n_k(1 - n_k)$ is maximal at $n_k = 1/2$ and vanishes for $n_k = 0,1$. Therefore, the same considerations discussed in Sec.~\ref{sec:mpemba} apply, leading to the conclusion that $\Delta(\braket{Q^2_A}_c)$ exhibits the Mpemba effect under the same conditions as $\Delta S^{(G)}_A$. This feature is not surprising, as the entanglement entropy and the even cumulants of the particle number operator $Q_A$ in Gaussian states are related through \cite{klich2009,song2012}
\begin{equation}
    S_A = \sum_{k=1}^{\infty} \frac{(2\pi)^{2k} |B_{2k}|}{(2k)!} \braket{Q_A^{2k}}_c,
\end{equation}
where $B_m$ are the Bernoulli numbers.
It is therefore expected that the cumulants themselves exhibit a similar behavior to $\Delta S^{(G)}_A$ in the symmetry restoration process. Hence we conclude that \eqref{eq:cumulant} (and higher cumulants) can be used as probes of symmetry restoration; this is indeed shown in Fig. \ref{fig:variance_of_charge}. 
This observation may prove useful for the experimental detection of the Mpemba effect, as the variance of the charge distribution is considerably easier to measure than Rényi entropies.

\section{Conclusions}
We have introduced an asymmetry measure for Gaussian states that quantifies the degree of symmetry breaking without leaving the Gaussian manifold. This Gaussian asymmetry is defined as the relative entropy between a Gaussian state and its symmetrised counterpart, which can be shown to minimise the relative entropy over the set of symmetric Gaussian states, thereby defining a monotone under symmetric Gaussian operations. 
This framework allows one to analyse symmetry restoration using only correlation-matrix techniques and to derive simple analytical predictions for the dynamics in the ballistic regime via the quasiparticle picture. 
Furthermore, we have shown that this notion of symmetrisation enables the use of charge fluctuations as a diagnostic of symmetry restoration. 

There are two natural directions for future work. First, it would be interesting to analyse more structured Gaussian protocols, including the presence of dissipation and measurements. While entanglement asymmetry has been partially investigated in such settings \cite{Caceffo_2024,digiulio2025,russotto2025dynamics}, the present quantifier is expected to provide a significantly simpler framework, enabling a more complete analytical understanding of symmetry restoration in these scenarios.
A second ambitious direction concerns the extension of the present ideas to integrable quenches in interacting systems. In this context, integrable states play a role analogous to Gaussian states in free systems: they form a closed set under integrable dynamics and, with few exceptions, are the only states that allow for analytical control. It would therefore be natural to explore whether a notion of integrability-preserving symmetrisation can be defined. However, this is considerably more subtle than in the free case, as there is no obvious or canonical way to associate a symmetric integrable partner to a given initial integrable state.

\section*{Acknowledgement}
We thank Colin Rylands, Bruno Bertini, Katja Klobas and especially Filiberto Ares for useful discussions on related topics.
Both authors acknowledge support from the European Research Council under the Advanced Grant no. 101199196 (MOSE).

\bibliographystyle{ytphys}
\bibliography{bibliography}
\end{document}